\documentclass[a4paper,10pt]{article}

\usepackage[english]{babel}
\usepackage{amsfonts}
\usepackage{amsmath}
\usepackage{amssymb}
\usepackage{graphicx}
\usepackage[a4paper]{geometry}%
\usepackage{indentfirst}
\providecommand{\U}[1]{\protect\rule{.1in}{.1in}}

\begin{document}

\pagenumbering{arabic}

\title{Finite-size and Fluctuation Effects on Phase Transition and Critical Phenomena using Mean-Field Approach Based on Renormalized $\phi^{4}$ Model:  I. Theory}

\author{ \bf R. M. Keumo Tsiaze $^{(1, 4)}$, S. E. Mkam Tchouobiap$^{(2)}$,  \bf A. J. Fotu\'{e}$^{(3)}$,\\ \bf C. Kenfack Sadem$^{(3)}$, J. E. Danga$^{(3)}$, C. Lukong Fa\"i$^{(3)}$ and  M. N. Hounkonnou*$^{(4)}$
\\
$^{(1)}$ Laboratoire de Science des Mat\'{e}riaux, Facult\'{e} des Sciences,
Universit\'{e} de Yaound\'{e} I,\\ B.P. 812 Yaound\'{e}, Cameroun.
\\
$^{(2)}$ Laboratory of Research on Advanced Materials and Nonlinear Science (LaRAMaNS),\\
Department of Physics, Faculty of Sciences, University of Buea, PO Box 63, Buea, Cameroon
\\
$^{(3)}$ Mesoscopic and multilayer structure laboratory, University of Dschang,\\ P.O.Box 479, Dschang, Cameroun.
\\
$^{(4)}$ International Chair in Mathematical Physics and Applications,\\
 University of Abomey-Calavi, 72 BP50 Cotonou, Republic of Benin.
\\
$^*$ Corresponding authors; {\textbf{email:} norbert.hounkonnou@cipma.uac.bj, keumoroger@yahoo.fr}\\}
%
\maketitle
\begin{abstract}
\large{
An  investigation of the spatial fluctuations and their manifestations in the vicinity of the quantum critical point within the framework of the renormalized $\phi^{4}$ theory is proposed.  Relevant features are reported through the Ginzburg-Landau-Wilson (GLW)-based calculations, combined with an efficient non perturbative technique.  Both the dimension and size, but also microscopic details of the system, leading to critical behavior, and strongly deviating from the classical mean-field approach far from the thermodynamic limit, are taken into account. Further, the important role that harmonic and anharmonic fluctuations  and finite-size effects can play in the determination of  the characteristic properties of corresponding various systems, involving phase transitions and critical phenomena, is then discussed in detail with emphasis on the qualitative validity of the analysis.}
\end{abstract}
\large{ \baselineskip 8mm \noindent \textbf{Keywords:}
\textit{Phase transition, critical phenomena, finite-size effects, harmonic and anharmonic fluctuations, Gaussian and non-Gaussian  approximation, Hartree-Fock decoupling, mean-field theory, $\phi^{4}$ model}}.\\
\\
\textbf{PACS numbers(s):} \textit{05.70.Fh, 64.60.an, 64.60.Ht, 05.70.Jk}

%


\section{Introduction}
\noindent

Nowadays the theory of phase transitions and critical phenomena seems to be well developed in general. It enables us to obtain both universal and non-universal properties for many model systems. However, there remains a number of unsolved important questions among which one can cite,  for instance, the role of fluctuations on the stabilization of the physical and electronic systems, the fluctuations that induced or prevent phase transitions, etc. There also exist   some  physical systems  having as a common characteristics the fact that the complex microscopic behavior underlies macroscopic effects.

In simple cases the microscopic fluctuations average out when larger scales are considered, and the averaged quantities satisfy classical continuum equation. Hydrodynamics is a standard example of this, where atomic fluctuations average out and the classical hydrodynamic equations emerge. Unfortunately, there is a much more difficult class of problems where fluctuations persist out to macroscopic wavelengths, and fluctuations on all intermediate length scales are important too.

In the last category are the problems of critical phenomena. The critical phenomena in the thermodynamic limit is characterized by the divergence of the correlation length $\xi$ near critical point. Nowadays, the experimental techniques have become so advanced that the correlation length $\xi$ can be pushed up to several thousand {\AA} and the samples under study become comparable with $\xi$. As a consequence, the effects of finite size of the samples on the critical phenomena become increasingly important. Generally such effects depend on the shape of the sample, the boundary condition, the dimension of the system and the number of components of the order parameter. In fully finite or quasi-one-dimensional systems the phase transition is smeared  out, whereas in thin films of thickness $L$ the critical temperature $T_{c}(L)$ is shifted with respect to the bulk $T_{c}$.

On the other hand, during the first half of the last century after the discovery of superconductivity the problem of fluctuation smearing of the superconducting transition was not even considered. In bulk samples of traditional superconductors the critical temperature $T_c$ sharply divides the superconducting and the normal phases. It is worth mentioning that such behavior of the physical characteristics of superconductors is in perfect agreement both with the Ginzburg-Landau (GL) phenomenological theory (1950) \cite{ginzburg} and the BCS microscopic theory of superconductivity (1957) \cite{bcs}. The characteristics of high temperature and organic  superconductors, low dimensional and amorphous superconducting systems studied today, strongly differ from those of the traditional superconductors discussed in textbooks. The transitions turn out to be much more smeared out. The appearance of superconducting fluctuations above the critical temperature leads to precursor effects of the superconducting phase occurring while the system is still in the normal phase, sometimes far from $T_c$. The conductivity, the heat capacity, the diamagnetic susceptibility, the sound attenuation, etc. may increase considerably in the vicinity of the transition temperature \cite{bennemann}.

    Recently, we proposed a microscopic renormalized Gaussian approach to critical fluctuations in the GLW model and finite-size scaling  to  describe  the phase behavior and critical phenomena in very varied systems~\cite{roger}. Within this more rigorous approach the effects of fluctuations are examined beyond the standard Gaussian approach (SGA)~\cite{stanley,ma} in more rigorous detail, and we are able to establish the insufficiencies of the mean-field theory (MFT), and also estimate the width of the critical region where corrections to MFT are important. The approach allowed us to obtain the effective functional of the Gaussian GLW Hamiltonian ($H_{GLW}$) expressed in terms of the collective variables without any procedure for further systematic improvement (e.g. by considering "higher order" terms).

 In this paper, we construct the $H_{GLW}$ (in the space of the fluctuating fields) devoted to the study of harmonic and anharmonic fluctuations and  their manifestations in the near-critical region for the GLW model. In the former case~\cite{roger} we took into account only the Gaussian approximation and showed that the obtained results were better and in good agreement with the experience as those found within the framework of the SGA method. Here we generalize the approach  taking into account the powers of field higher than the first one. We need to construct a renormalized GLW which includes the quartic term that also takes into account the microscopic details of the systems. The importance and role of the $\phi^{4}$ interaction were emphasized by Fisher and Wilson~\cite{wilson1,fisher1,fisher2,wilson2}. Its presence allows a low-temperature behavior and its absence leads to the divergence of the Hamiltonian. The main ingredient in the analysis is as follows:  the Fourier transform of the $\phi^{6}$ field theory can be seen as an interaction between the Fourier components of the order parameter, which are in fact a combination of harmonic and anharmonic fluctuations modes of the order parameter.  This makes  it  possible  for  the  coefficients  of  the  $\phi^{2}$  and   $\phi^{4}$ terms to  be  renormalized  by  the  $\phi^{6}$ term coefficient. The approach is a strategy for dealing with problems involving many length scales. The  strategy is to tackle the problem in steps, one step for each length scale. In the case of critical phenomena, the problem is,  technically,  to carry out statistical averages over thermal fluctuations on all size scales. The method is to integrate out the fluctuations in sequence, starting with fluctuations on an atomic scale and then moving to successively larger scales until fluctuations on all scales have been averaged out. The integration of this deviation in the self-consistency on the entire spectrum of the lattice vibration, conducted to a very improved self consistent energy. This improvement of the self consistent problem by the processed harmonic and anharmonic fluctuations goes, for instance, to reflect itself on the thermodynamic quantities and electronic parameters, in the neighborhood of the phase transition, in particular the singularities as function of the control parameter, resulting notably in a substantial modification of the critical point and thermodynamic quantities. The resulting self consistent problem of this approach is solved analytically, what  permits us to extract an effective theory, with notably a mean-field critical temperature renormalized by anharmonic fluctuations.
The present work does not present any renormalization based method in the sense of a renormalization group theory approach but rather a self-consistent method (SCM) improving on Landau theory (LT). Apart from calculations of exponents and scaling functions, it is necessary to develop techniques for obtaining the corrections to the asymptotic critical behavior, in terms of a small number of non universal parameters which can be fit by experimental results on different materials. In this way, it is hoped that a more rigorous confrontation between experiments and theory can be achieved.
%
%
\section{Formulation}
%
 The starting point of our investigation is the following $H_{GLW}$ functional,
\begin{eqnarray}
H_{\textrm{GLW}} \simeq \int\frac{d^dr}{\xi_0} \Big[(\nabla_r\phi(\textbf{r}))^2 + F_{\textrm{L}} \Big],
\end{eqnarray}
where $F_{\textrm{L}}$ is the Landau free energy given by
\begin{eqnarray}
F_{\textrm{L}} = a(T){\phi}^2(\textbf{r})  + b(T_c){\phi}^4(\textbf{r}) +  u_{0}{\phi}^6(\textbf{r}) + \cdots.
\end{eqnarray}
Here, $\phi$ is the model order parameter characterizing the mode displacement. The quadratic coefficient $a(T) = a_{0}(T - T_{c})$ vanishes linearly as the temperature approaches the mean-field critical temperature $T_{c}$. For stability purposes, the coefficient constant $u_0$ is chosen such that $u_0 > 0$. The LT of second order (ferromagnetic) phase transitions, for example, amounts to postulating the existence of a development of type $b(T_c) = \frac{k_BT^3_c}{12T^2}\Big\vert_{T = T_c}$ close to $T_c$, what is not right for the superconductors where the microscopic relations between the GL parameters are such as $\frac{a_0^2}{b(T_c)} = \frac{8\pi^2}{7\zeta(3)}\nu,$ where $\zeta(x)$ is the Riemann zeta-function ($\zeta(3)$ = 1.202...), $k_B = 1.38\times 10^{-23}$ W s/K is Boltzmann's constant and $\nu$ a parameter which will be defined later. But just like the quadratic coefficient, the constant $b(T_c)$ must be to improve to take into account both the microscopic details, dimension and size of the system. The real expressions of $a(T)$ and $b(T)$ within the framework of a renormalized theory will be defined later. $\xi_0$ is the coherence length of the sample (interpolated down to T = 0 K). In superconductors, average extension $\xi_0$ of a Cooper pair as a measure of the distance within which the correlation forming Cooper pairs is active. In magnetic systems, $\xi_0$ represents the lowest length (microscopic) for spin correlations (spin waves), while in solids, it represents that of the vibrational or phonon modes.  It is in general of the order of the interaction range. However, fluctuations with wavelength $\xi_0$ will be seen to be always negligible.  $d$ is the dimensionality of physical space.

The polynomial in Eq. (2) originates from a power series expansion of some potential  $V(\phi)$ \cite{landaulifshitz, grundland}. The order to  which its terms are kept depends on physical considerations  given by renormalization theory \cite{amit} and on the type of symmetry-breaking  effect \cite{galam} Eq. (2) is supposed to model. The odd powers in Eq. (2) are dropped as a result  of e.g. time-reversal invariance. A "$\phi^{6}$" model is of interest since, on contrast to "$\phi^{4}$" models  it can  describe both second ($b(T) > 0$) and first order ($b(T) <0$) phase transitions. Moreover, it  displays a "butterfly" catastrophe \cite{poston} which is more complex than the cusp catastrophe of the "$\phi^{4}$" model. The "$\phi^{6}$" model allows the ground state to be up to triply degenerate; situations with no degeneracy, double degeneracy or triple degeneracy can be studied by varying parameters in the expansion. The case $a(T)  = b(T) = 0$ is of particular interest in the context of tricritical points on phase diagrams \cite{aharony, gordon}.

LT \cite{landaulifshitz} implicitly assumes that analyticity is maintained as all space-dependent fluctuations are averaged out. The loss of analyticity arises only when averaging over the values of the overall average (order parameter) $\phi$. It is this overall averaging, over $\exp(-\beta F_{L})$, which leads to the rule that $F_{L}$ must be minimized over $\phi$. At temperatures that are far below, or far above a critical point, the behavior of the order parameter reassembles a tranquil ocean with no significant amount of thermal noise in its fluctuations. But fluctuations become increasingly important near the critical point as the correlation length diverges. At the second-order phase transition, infinitely long-range "critical fluctuations" develop in the order parameter. The study of these fluctuations requires that we go beyond LT.

The SGA to the problem posed by the $H_{GLW}$ is to decompose $\phi(\textbf{r})$ into its Fourier components $\phi(\textbf{q})$ according to
\begin{equation}
\phi(\textbf{r}) = L^{-d/2}\sum_{|q|< \Lambda}\phi(\textbf{q})\phantom{.}e^{i\textbf{q}\cdot{\textbf{r}}}.
\end{equation}
According to Eq. (3) the limit on wavelengths means that the integration over $q$ is restricted to values of $q$ with $\vert q \vert < \Lambda$. Averaging over long-wavelength fluctuations now reduce to integrating over the variables $\phi(q)$, for all  $\vert q \vert < \Lambda$ \cite{wilson3}. There are many such variables; normally this would lead to many coupled integrals to carry out, a hopeless task. Considerable simplifications will be made below in order to carry out these integrations.

We need an integrand for these integrations. The integrand is a constrained sum of the Boltzmann factor $k_{B}$ over all atomic configurations. The constraints are that all $\vert q \vert < \Lambda$ are held fixed. this is a generalization of the constrained sum in the LT. We shall assume Landau's analysis is still valid for the form of $H_{\textrm{GLW}}$, that is, $H_{\textrm{GLW}}$ is given by Eq. (1). However, the importance of long-wavelength fluctuations means that the parameters $a(T)$ and $b(T)$ depend on $\Lambda$ and then on the dimension $d$ of the system. The $d$ dependence of $a(T)$ and $b(T)$ will be determined shortly. However the breakdown of analyticity at the critical point is a simple consequence of this $d$ dependence. Details will also be discussed shortly.

The change of variables to the Fourier modes, the $H_{\textrm{GLW}}$ expansion is given by the expression
\begin{eqnarray}
&& H_{GLW} \simeq  \sum_{q<\Lambda}{\bar{G}}^{-1}(\textbf{q})\phi(\textbf{q})\phi(-\textbf{q})    \nonumber\\
&& + \frac{b(T_{c})}{L^{d}}\sum_{\{q\} < \Lambda}\phi(\textbf{q}')\phi(\textbf{q}'')\phi(\textbf{q}''')\phi(-\textbf{q}'-\textbf{q}''-\textbf{q}''') \nonumber\\
&& + \frac{u_{0}}{L^{2d}}\sum_{\{q_{i}\} < \Lambda}\phi(\textbf{q}_{1})\phi(\textbf{q}_{2})\phi(\textbf{q}_{3})\phi(\textbf{q}_{4})\phi(\textbf{q}_{5})\phi(-\textbf{q}_{1}-\textbf{q}_{2}-\textbf{q}_{3}-\textbf{q}_{4}-\textbf{q}_{5}) \\
 && + \cdots  \nonumber
\end{eqnarray}
where the $q$-mode function $\bar{G}^{-1}(\textbf{q}) = a(T) + cq^{2}$, and $L$ is the linear dimension of the sample. Therefore, it is useful to express the partition function $\mathcal{Z}$ as a functional integral of the wave vector fluctuations $\phi(\textbf{q})$. Accordingly, $\mathcal{Z}$ generalizes and therefore factorizes into
\begin{eqnarray}
&& \mathcal{Z} = \int D\phi\cdot \exp\Bigg[ -\beta \Bigg\{\sum_{q}{\bar{G}}^{-1}(\textbf{q})\phi(\textbf{q})\phi(-\textbf{q}) \nonumber\\
&& +\frac{b}{L^{d}}\sum_{\{q\} < \Lambda}\phi(\textbf{q}')\phi(\textbf{q}'')\phi(\textbf{q}''')\phi(-\textbf{q}'-\textbf{q}''-\textbf{q}''')  \nonumber\\
&& + \frac{u_{0}}{L^{2d}}\sum_{\{q_{i}\} < \Lambda}\phi(\textbf{q}_{1})\phi(\textbf{q}_{2})\phi(\textbf{q}_{3})\phi(\textbf{q}_{4})\phi(\textbf{q}_{5})\phi(-\textbf{q}_{1}-\textbf{q}_{2}-\textbf{q}_{3}-\textbf{q}_{4}-\textbf{q}_{5})  + \cdots\Bigg\}\Bigg],\cr
&&
\end{eqnarray}
where $D\phi$ is given by
\begin{equation}
\mathcal{D}\phi = \prod_{q}(2\pi)^{-1}d\phi(\textbf{q})d\phi(-\textbf{q}),
\end{equation}
and $\beta = 1/k_{B}T$. $\mathcal{D}\phi$ is used to denote the measure of the functional integral. This quantity ensures that the total probability is normalized to unity through a constant of proportionality $1/(2\pi)$\cite{amit, reichl}. $(2\pi)^{-1}$ left out is formally divergent in the thermodynamic limit;  it does not affect averages that are obtained from derivatives of such integrals.

In papers \cite{tuszyinski2} and \cite{tuszyinski3}, a yet another approach to critical fluctuations has been proposed. It is based on the fact that quartic term is dominant and near criticality since $ a(T) \rightarrow 0$ as $T \rightarrow T_{c}$. In the first paper \cite{tuszyinski2} using mean-field approximation (MFA), the influence of homogeneous fluctuations was examined through the expansion
\begin{equation}
H_{\textrm{MFA}} \cong H_{\textrm{GLW}}(\bar{\phi}) + \lambda_{2}(\phi - \bar{\phi})^{2} + \lambda_{4}(\phi - \bar{\phi})^{4} + \cdots
\end{equation}
where $\bar{\phi}$ is the equilibrium value of the order parameter and $H_{\textrm{GLW}}$ is given by Eq. (1) without the Ginzburg term.  In the second paper \cite{tuszyinski3} expansion like that in Eq. (7) was studied for case other than spontaneous second-order transitions, i.e., for field-induced transitions, first-order transitions, and liquid-vapor transitions. In both papers calculations were performed using non-Gaussian integral given by\cite{witschel, gradshteyn}
\begin{eqnarray}
&&\int_{0}^{\infty}\phi^{2mp-1} \exp{\left (-\lambda_2\phi^{2m} - \lambda_4\phi^{4m}\right)} \ d\phi \nonumber\\
&&= (2m)^{-1}(2\lambda_4)^{-p/2}\Gamma(p) D_{-p}\Big(\lambda_2/\sqrt{2\lambda_4}\Big) \exp{\Big(\lambda_2^2/8\lambda_4\Big)},
\end{eqnarray}
and analyzed in all regimes including finite-size. $D_{-p}$ is the parabolic cylinder function \cite{abramowitz} and $\Gamma$ is usually so-called gamma function. The most interesting conclusions were that Gaussian approximation fails for all values of parameters except for in the thermodynamic limit ($V \longrightarrow \infty$). As a result, power-law predictions of the Gaussian prediction are incorrect, except for at $V = \infty,$ and should therefore be replaced by exponential asymptotic behavior as predicted by non-Gaussian methods. This results, for example, in vastly different asymptotic predictions for finite-size scaling as was discussed at length in both papers \cite{tuszyinski2,tuszyinski3}.  These two papers, however, dealt exclusively with mean-field properties of non-Gaussian critical fluctuations, setting the Ginzburg term to zero. This,  of course, neglected a very important property of critical systems, i.e., their spatial inhomogeneity \cite{ginzburg}. Nonetheless, interesting results were obtained \cite{tuszyinski2,tuszyinski3}, leading asymptotically to the Gaussian approximation results and also providing finite-size scaling for $T \neq T_{c}$. Therefore, we believe that the proposed non-Gaussian method offers large region of analyticity and possesses better convergence properties. The present paper is intended to provide another insight into the problem by using non-Gaussian way of averages  calculation and keeping a significant part of the Ginzburg term. We believe the renormalized $H_{GLW}$ is capable of properly describing both the critical and near-critical regimes. A non-Gaussian method of calculation, however, must be employed to adequately reveal the deviation from asymptotic behavior and thus the crossover phenomena.

%
\subsection{Non-Gaussian Fluctuations}
%

A major limitation of the GLW theory is its incapacity to account for the characteristics of the intermediate mode between the adiabatic mode on the one hand and the non-adiabatic mode on the other hand. It is also and especially its incapacity to take account of the microscopic details and the dimensionality of the systems. For instance,  the GLW theory also
predicts incorrect results, like an unphysical phase transition in 1D, or incorrect critical indices, in higher dimensions.  However, Scalapino, Sears and Ferrell \cite{scalapino} showed that, at least in 1D, this failure is due to an improper treatment of fluctuations, so the complete success of the GLW theory will depend essentially on how fluctuations are taking into account. Both mean-field theories and Gaussian-based power expansions as well as renormalisation-group calculations are concerned with the asymptotic properties of critical systems in the sense of infinite size and immediate proximity to $T_c$. Thus they predict singular power-law behavior of the systems with universal exponents and scaling functions. As has been recently made abundantly clear, the asymptotic regime is indeed very small and agreement with experimental data deteriorates very rapidly outside the near-critical region; in fact extrapolation from the critical-state equations fails outside criticality and vice-versa, and the analytic noncritical equations of state do not reproduce the correct singular behavior at criticality. Thus there is a real challenge and a need to develop a method of calculation that both incorporates the asymptotic critical behavior and the crossover to regular regime.

In the approximation that we propose,  we wish to include those mode-mode coupling terms which involve balanced pairs of $q $ and -$q$ wave vectors, assuming that the remaining combinations are less important as they may lead to numerous cancelations. We can analytically understand the effective free energy of the exponential argument in Eq. (5) as that containing the quadratic term for the modes of free fluctuations of the order parameter $\phi(q)$ (quadratic form), the quartic term for the modes of harmonic fluctuations of the order parameter $\phi(q)$, and the sixth-order term as the anharmonic interactions between these modes. This condition seems to be effectively justified if the sixth-order coefficient of the expansion is small.

Accordingly, it is important in this case to precise the following scenario:

(i) The $\phi^{2}$ and $\phi^{4}$ terms are linked to the essential fluctuations.
(ii) The $\phi^{6}$ term is linked to the redundant fluctuations.
(iii) All other terms correspond to the unessential fluctuations, i.e. contain no essential new physics and in fact are "irrelevant" (to the zero-temperature critical behavior in the sense described by Wilson \cite{ma, wilson1, fisher1}).

Therefore, the  sixth-order term appears as and seems to play the role of interaction terms between the Fourier components of the order parameter, which are in fact the combination of harmonic and anharmonic fluctuation modes of the order parameter. The sixth-order term can thus contribute to improve the  quadratic and quartic coefficients. This makes  it  possible  for  the  coefficients  of the  $\phi^{2}$ term to  be  renormalized "an-harmonically"  and the  $\phi^{4}$ term to be  renormalized "harmonically" by  the  $\phi^{6}$ term coefficient. It should keep in mind that the GLW theory is a phenomenological theory, founded on the intuition, with its own laws and rules. It accounts for the phenomena and it is that its justification.

In this study, we will go beyond the GLW theory in order to evaluate the influence of the fluctuations on the critical phenomena and particularly critical temperature for systems with short-range interactions. We use a Hartree-Fock decoupling for $\phi^{6}$ interactions in the consideration of the absence of long-range interactions. This approximation will consist in considering that the Fourier components interact (in harmonical and anharmonical manner) only through the mean field produced by other modes. In the case of continuous phase transition this idea allows us to decouple the $\phi^{6}$ term into a sum of product of two quantities with even powers:
\begin{eqnarray}
&&\frac{1}{L^{2d}}\sum_{\{q_{i}\} < \Lambda}\phi(\textbf{q}_{1})\phi(\textbf{q}_{2})\phi(\textbf{q}_{3})\phi(\textbf{q}_{4})\phi(\textbf{q}_{5})\phi(-\textbf{q}_{1}-\textbf{q}_{2}-\textbf{q}_{3}-\textbf{q}_{4}-\textbf{q}_{5}) \nonumber\\
&&\approx 15 \sum_{q} \Bigg(\frac{1}{L^{2d}}\sum_{\{q_{i}\} < \Lambda}\Big\langle\big|\phi(\textbf{q}')\phi(\textbf{q}'')\phi(\textbf{q}''')\phi(-\textbf{q}'-\textbf{q}'' -\textbf{q}'''))\big|\Big\rangle\Bigg)\phi(\textbf{q})\phi(-\textbf{q})
\nonumber\\
&& + \frac{15}{L^{d}}\sum_{\{q_{i}\}} \Bigg(\frac{1}{L^{d}}\sum_{q < \Lambda} \Big\langle\big|\phi(\textbf{q})\phi(-\textbf{q})\big|\Big\rangle\Bigg)\phi(\textbf{q}')\phi(\textbf{q}'')\phi(\textbf{q}''')\phi(-\textbf{q}' -\textbf{q}''-\textbf{q}''')) + \cdots.
\end{eqnarray}
The factor 15 takes into account all possible contractions(statistical average of two or four modes among six). The terms between brackets act like mean fields. The sixth order term thus uncoupled consequently becomes a combination of quartic and quadratic terms, enabling us to obtain an effective non-Gaussian theory. With all these important preliminaries and considerations, one may then conveniently write the effective $H_{\textrm{GLW}}$ in the absence of an external field as
\begin{equation}
H_{\textrm{eff}}[\phi] = \int \frac{d^dr}{\xi_0}\bigg[\big(\nabla_r\phi(\textbf{r})\big)^2 + a^{\ast}(T)\phi^2(\textbf{r})+ b^{\ast}(T))\phi^4(\textbf{r}) + \cdots \bigg],
\end{equation}
and the result of applying Eq. (3) to Eq. (10) is
\begin{eqnarray}
&& H_{\textrm{eff}}[\phi] \simeq  \sum_{q<\Lambda}{\bar{G^{\ast}}}^{-1}(\textbf{q})\phi(\textbf{q})\phi(-\textbf{q})    \nonumber\\
&& + \frac{b^{\ast}(T)}{L^{d}}\sum_{\{q\} < \Lambda}\phi(\textbf{q}')\phi(\textbf{q}'')\phi(\textbf{q}''')\phi(-\textbf{q}'-\textbf{q}''-\textbf{q}''') + \cdots
\end{eqnarray}
where the renormalized $q$-mode function $\bar{G^{\ast}}^{-1}(\textbf{q}) = a^{\ast}(T) + cq^{2}$.

The quadratic coefficient $a^{\ast}(T)$ and the quartic coefficient $b^{\ast}(T)$ are now respectively given by
\begin{equation}
a^{\ast}(T) = a(T) + \Omega_d(T)
\end{equation}
\begin{equation}
b^{\ast}(T) = b(T_{c}) + \Theta_d(T).
\end{equation}
The term
\begin{equation}
\Omega_d(T) = \frac{15u_{0}}{L^{2d}}\sum_{\{q_{i}\} < \Lambda}\Big\langle\big|\phi(\textbf{q}')\phi(\textbf{q}'')\phi(\textbf{q}''')\phi(-\textbf{q}'-\textbf{q}'' -\textbf{q}''')\big|\Big\rangle,
\end{equation}
includes a new contribution to the critical point and corresponds to the anharmonic variance of the order parameter at a single point in space evaluated at temperature T. This term competes with Landau quadratic coefficient $a(T)$. Therefore it determines the critical line and hence incorporates both the asymptotic critical behavior and the crossover to the regular regime.

The renormalized quartic coefficient obtained within the renormalized GLW approach is now given by Eq. (13) with the correction term $\Theta_d(T)$.  The term
\begin{equation}
\Theta_d(T) = \frac{15u_{0}}{L^{d}}\sum_{q} \Big\langle\big|\phi(\textbf{q})\phi(-\textbf{q})\big|\Big\rangle,
\end{equation}
then also includes another new contribution to the quartic coefficient. This term competes with Landau quartic coefficient $b(T_c)$ and defines the tricritical crossover exponent. Both scaling densities $\Omega_d(T)$ and $\Theta_d(T)$ are relevant, i.e., exhibit critical fluctuations. Now the quartic coefficient can be cancelled at some point highlighting the existence of a tricritical point. This fact represents the principal difference between the transitions corresponding to the tricritical Gaussian fixed point and the ordinary second-order Gaussian fixed point. Hence the tricritical fixed point can be characterized as the simultaneous instability of the system to two types of critical fluctuations. Correction to the molecular-field tricritical behavior due to critical fluctuations will be discussed in a forthcoming paper. It is found that the asymptotic tricritical form of some thermodynamic quantities or functions is not a power law but a power law multiplied by a fractional power of a logarithm.

Taking a look at the influence of spatial fluctuations on the stability of the renormalized GLW analysis, the importance of fluctuations seems to be evident and the divergence of the critical coefficient of the second-order term $a^{\ast}(T)$ is now related to the system size $L$ and dimensionality $d$ of the system. Therefore it is useful to determine the temperature at the critical point where the divergence of quadratic coefficient should be observed.  Accordingly, the modified critical temperature is given by
\begin{equation}\label{eq15}
T^{\ast}_c = T_c - \frac{{\Omega_d}_c}{a_0}
\end{equation}
where ${\Omega_d}_c = \Omega_d(T^{\ast}_c)$ is the solution of the self-consistent Eq. (25) that will be discussed in the next subsection.

Eq. (16) shows that $T^{\ast}_{c}$ can be neglected if the anharmonic fluctuations are too high. However, taking into consideration the $\phi^{6}$ term contained in the $H_{\textrm{GLW}}$ functional leads to the emergence of the limitations of the GLW approach and MFA related to the critical temperature $T_{c}$, which is presented here as a characteristic scale of temperature related to both thermic fluctuations and finite-size effects rather than the transition temperature \cite{roger}. Although that limitation is a disadvantage to the exactness of the GLW approach or MFA, its predictions are not at all lacking interest and only the importance of the anharmonic term $\Omega_{dc}$  should determine its degree of validity.

Structural phase transition (for example) is accompanied by a change in structure. Some of these changes in structure occur without macroscopic diffusion of matter in solids. They are initiated by local motions of atoms or molecular groups which can distort the lattice in the high temperature phase to form structures of lower symmetry at lower temperatures. These movements around equilibrium positions do not occur instantaneously at temperature $T_{c}$, and they are actually initiated at a temperature lower or higher than the transition temperature \cite{ppapon}. In a way, this pretransition or pretransformation phenomenon is the equivalent of the nucleation process. Eq. (16) can thus contribute to clarify this viewpoint, since $\Omega_{dc}$ which is the solution of a $n$ degree  polynomial can be a positive or negative quantity.
%
%
\subsection{Mean-field approximation of $\Omega_d(T)$, $\Theta_d(T)$ and thermodynamic quantities}
%
A correct treatment of $\Omega_d(T)$ shows that
$\Theta_d(T)$
 is much more complex. Once critical fluctuations are not treated as the constants, one could imagine expanding order parameter in a Taylor's series about its value at some central location $\textbf{r}_{0}$. This means that $H_{\textrm{GLW}}$ could be a complicated functional of $\phi$, an expression that  is hard to write down, with several parameters, instead of the simple GLW form with only two parameters $a(T)$ and $b(T_c)$. Consequently, a natural recourse is to use Gaussian approximation.

Gaussian measures play a central role in many fields: in probability theory as a consequence of the central limit theorem, in quantum mechanics, in quantum field theory, in the theory of phase transitions in statistical physics. In this section, calculations will be performed and analyzed in all regimes including finite sizes, using the Gaussian integrals in the form \cite{abramowitz},
\begin{eqnarray}
&& \int_{0}^{\infty}y^{2n}\exp(-py^{2})dy = \frac{(2n-1)!!}{2(2p)^{n}}\sqrt{\frac{\pi}{p}} \nonumber\\
&& p > 0, \phantom{...} n = 0, 1, 2, \cdots; \phantom{...}(2n-1)!! = 1\cdot3\cdot5\cdots(2n-1).
\end{eqnarray}
We can formulate the Hamiltonian by averaging the new contributions. The procedure for identifying the variational parameter of the quadratic and quartic coefficients, $\Omega_d(T)$ and $\Theta_d(T)$, and for determining the value of  $T^{\ast}_c$ and thermodynamics quantities is as follows. First the Hamiltonian is derived in the MFA and the obtained $\Omega_d(T)$ is substituted for the self-consistent equation. The derived equation is used to determine both the value of $T^{\ast}_c$ and thermodynamic quantities.

As a result, it is shown that the value of $T^{\ast}_c$ and thermodynamic quantities strongly deviate from the classical MFA but thermodynamic quantities adopt mean-field critical exponents provided by the classical Landau approach. We find that the variational parameter of the Hamiltonian obtained by renormalizing that of the quadratic term, corresponding to the square of angular frequency, is proportional to $T^{\ast}_c - T$.

The thermal average of a physical  quantity $y(\phi)$ is given by
\begin{equation}
\langle{y}\rangle  = \int_{0}^{\infty}y(\phi)\exp[-\beta{H_{\textrm{eff}}}]d\phi\Bigg/\int_{0}^{\infty}\exp[-\beta{H_{\textrm{eff}}}]d\phi.
\end{equation}

The obvious difficulty that the transformation given by Eq. (3) brings relates to the mode-mode coupling present in the last term in Eq. (11). Consequently, a natural recourse is to use the Gaussian approximation where $H_{\textrm{eff}}[\phi]$ is truncated to
\begin{equation}
 H_{0}[\phi] =  \sum_{q<\Lambda}{\bar{G^{\ast}}}^{-1}(\textbf{q})\phi(\textbf{q})\phi(-\textbf{q})
\end{equation}
This then conveniently factorizes the partition function as
\begin{equation}
\mathcal{Z}_{0} = \prod_{q < \Lambda}\sqrt{\frac{\pi}{\beta\Big(a_0(T - T_c) + \Omega_d + q^2 \Big)}}.
\end{equation}
By making a transition to continuum through
\begin{equation}
\sum_{q}(...) = L^d(2\pi)^{-d}\int_{-\infty}^{\infty}(...)d^dq,
\end{equation}
the free energy is
\begin{equation}
F_{0} = -\frac{k_{B}TL^{d}}{2(2\pi)^{d}}\int_{0}^{\infty}\ln\Bigg[\frac{\pi}{\beta\Big(a_0(T - T_c) + \Omega_d + q^2 \Big)}\Bigg]d^dq
\end{equation}
The heat capacity of the system is given by
\begin{equation}
C_{0} \cong - T\frac{\partial^2 F_{0}}{\partial T^2} = \kappa_d \Big[\frac{T}{T_c}\Big]^2\Big(a_0 + \frac{\partial\Omega_d}{\partial T}\Big)^2\Big(\epsilon + \frac{\Omega_d}{a_0 T_c}\Big)^{d/2 - 2} + \textrm{less singular terms},
\end{equation}
where
\begin{equation}
\kappa_d = \frac{k_B L^d (\xi_0)^{-d}}{2^{d - 1}\pi^{d/2}a^2_0\Gamma(d/2)}\int_0^{+\infty}\frac{x^{d-1}dx}{(1+x^2)^2}, \phantom{...} \epsilon = \frac{T - T_{c}}{T_{c}}.
\end{equation}
The element volume is $\Re_{d}q^{d-1}dq$, with $\Re_{d} = \frac{2\pi^{d/2}}{\Gamma(d/2)}$ the sphere unit surface in $\mathbb{R}^{d}$ space. $\kappa_d$ is an integral correction constant which is lattice dimension dependent. The behavior of the integral correction constant changes dramatically at $d = 4$. For $d > 4$ the integral diverges at large $x$ and is dominated by the upper cut-off $\Lambda$, while for $d < 4$, the integral is convergent in both limits. Although the expression of the specific heat is presented in not so complicated form, it is not quite transparent to know what type of behavior it will exhibit depending on dimension. Only the nature of $\Omega_d$ will make it possible to describe its behavior.

The dominant behavior of $C_{0}$ close to $T^{\ast}_c$ is through $C^{\ast}_{0} \sim |\epsilon^{\ast}|^{d\nu - 2}$, where the reduced and renormalized temperature $\epsilon^{\ast} = \epsilon + \frac{\Omega_{dc}}{a_0 T_c}$ and $\nu = 1/2$, so that we obtain $\alpha = 2 - d\nu$ as in the SGA . When thermal fluctuations can be neglected, as it appears to be the case in conventional superconductors, the specific heat exhibits at the transition temperature $T_{c} (L\rightarrow \infty)$  a step discontinuity. This differs drastically from the behavior when thermal fluctuations dominate. Due to the finite-size effect,  the specific heat peak occurs at a temperature $T^{\ast}_{c}(L)$ shifted from the homogeneous system by an amount proportional to $L^{-1/\nu}$, and the magnitude of the peak located at temperature $T^{\ast}_{c}(L)$ scales as $L^{\alpha/\nu}$ \cite{privman1}. This result seems to be qualitatively in good agreement with the well-known result since $C_{0} \sim |T - T^{\star}_c|^{d\nu - 2}$ \cite{ma}.

%
%
\subsubsection{Expressions of fluctuating quantities $\Omega_d(T)$, $\Theta_d(T)$}
%

The structural phase transition has been, hiterto, treated as follow\cite{katsuhiko}: In the "quasiharmonic" approximation \cite{kwok, axe}, it was assumed that harmonic frequency for the unstable mode is purely imaginary, namely, $\omega^{2}_{0} < 0$ under the short- and long-range force constant competing with each other. Then the contribution from the anharmonic term was derived self-consistently, which is proportional to T. As a result, the square of the soft-mode frequency was given by $\omega^{2} = \omega^{2}_{0} + c\langle u^{2}\rangle\propto T - T_{c}$ where $u$ denote the displacement from the interatomic distance. Other soft-mode theories are LT and MFT \cite{cowley, fleury}. In the expansion of free energy, the normal coordinate was adopted as the order parameter. It was shown that the critical coefficient of the second-order term, $a(T)$, equals the square of the soft-mode frequency, $a(T) = \omega^{2}\propto T_{c} - T$. The LT gives a qualitatively correct view of the soft mode, but cannot explain how it occurs; furthermore, this theory is not valid close to $T_{c}$ where critical fluctuations can no longer be neglected. So its complete success will depend essentially on how fluctuations are taken into account.

In the present work, we show the approach to this problem from the microscopic point of view, both in terms of the order parameter and self-correlation functions $\langle\phi^{2}\rangle$ and $\langle\phi^{4}\rangle$. From Eqs. (17), (18) and (21), we obtain the result that the temperature dependence of the variational parameters, $\Omega_d(T)$ and $\Theta_d(T)$, corresponding to the anharmonic and harmonic fluctuations of the order parameter, are presented by
\begin{equation}
\Omega_d = \frac{K_d}{V}\bigg[\frac{T}{T_c}\bigg]^{2}\bigg(\epsilon + \frac{\Omega_d}{a_0 T_c}  \bigg)^{\frac{d}{2} - 2} \textrm{where} \phantom{..} V = L^{d},
\end{equation}
and
\begin{equation}
\Theta_d = \mathcal{K}_d\bigg[\frac{T}{T_c}\bigg]\bigg(\epsilon + \frac{\Omega_d}{a_0 T_c}\bigg)^{\frac{d}{2} - 1}.
\end{equation}
$K_d$ and $\mathcal{K}_d$ are the dimension-dependent constants,  respectively, and explicitly established as follows:
\begin{equation}
K_d = \frac{45u_{0}{k_B}^{2}\xi_0^{-d}}{2^{d - 1}\pi^{d/2}a^2_0\Gamma(d/2)}\int_0^{+\infty}\frac{x^{d-1}dx}{(1+x^2)^2},
\end{equation}
and
\begin{equation}
\mathcal{K}_d = \frac{15u_{0}{k_B}\xi_0^{-d}}{2^{d - 1}\pi^{d/2}a_0\Gamma(d/2)}\int_0^{+\infty}\frac{x^{d-1}dx}{(1+x^2)}.
\end{equation}
The equations ((25) and (26)) are derived taking into account the expected fact that  $\Omega_d(T)$ and $\Theta_d(T)$ do not explicitly depend on wave number $q$. The behavior of $K_d$ is the same one as that of $\kappa_d$. For certain values of $d,$ the integral $\mathcal{K}_d$ diverges at large $x$. This is not a first time when such problem  arises and we know how to deal with it: this ultra-violet (UV) divergence is related to the restrictions on the applicability of the GLW functional for $|q| \gtrsim \xi^{-1}$, so the integral has to be cut off at $\xi\cdot{|q|} = x_c \sim 1$.

It is well established that all correlation functions (in dimension $d < 4$) have a large cut-off limit after a simple renormalization, that is, after one has taken the deviation from the critical temperature as a parameter. The field amplitude renormalizations are finite. In the critical phenomena situation instead, the fluctuations, which is related to microscopic parameters of the theory, is fixed. This means that, after the introduction by rescaling of the cut-off $\Lambda$, quadratic coefficient remains finite when $\Lambda \rightarrow \infty$ for $T \neq T^{\ast}_{c}$.

As we are concerned with phase fluctuations, Eqs. (25) and (26) show that their effect is dimension $d$ dependent. It has been demonstrated that the inclusion of phase fluctuations leads to a reduction in the degree of order in $d > 2$ and to its complete destruction on $d \leq$ 2; for $d > 2$, the phase fluctuations are finite while they become asymptotically large for $d \leq$ 2 \cite{stanley, ma, brezin}. The more general result has been established and known as the \textit{Mermin-Wagner theorem} \cite{mermin}: it states that there is no spontaneous breaking of a continuous symmetry in systems  with short-range interaction in dimension $d \leq$ 2, and as a corollary the borderline dimensionality of 2, known as the \textit{lower critical dimension} $d_l$ has to be treated carefully. With regards to Eq. (26) dimension $d = 2$  seems to play a crucial role in agreement with the \textit{Mermin-Wagner theorem}.  Indeed, when $d = 2$, the quantity $d/2 - 1$ vanishes and the self-consistent equation for the harmonic correction term $\Theta_d$ (Eq. (26)) is reduced to a linear function of T with a positive coefficient as $\Theta_d = \mathcal{K}_d\Big[\frac{T}{T_c}\Big]$. On the other hand, above d = 4, the MFT predicts correctly the universal quantities, whereas it is definitely not valid for dimension 4 and below. (Eq. (25) seems to arise this aspect with the quantity $d/2 - 2$ which is canceled when $d = 4$, and the self-consistent equation for the anharmonic correction term $\Omega_d$ (Eq. (25)) is reduced to a function of $T^{2}$ with a positive coefficient as $\Omega_d = K_d\Big[\frac{T}{T_c}\Big]^{2}$. Although the expressions of the correction terms are presented in not so complicated form, it is not quite transparent to know what type of behavior they will exhibit depending on dimension, temperature and the size of the system. The possibility that $\Theta_d$ or $\Omega_d$ increases with decreasing temperature could highlight the quantum character of the system according to both the dimensionality and the size. However, the numerical aspect which will be approached in another work, will enable us to better clarify the behavior of $\Theta_d$ and $\Omega_d$ and to give them a suitable physical direction.

The renormalized critical coefficient of the second-order term $a^{\ast}(T)$, equals the square of the renormalized soft-mode frequency, $a^{\ast}(T) = {\omega^{\ast}}^{2}\propto T^{\ast}_{c} - T$, but the critical temperature now takes into account both the dimension of the system, its size and critical fluctuation.

%
%
\subsubsection{Thermodynamic quantities}
%

To determine the degree of validity of this approach, let us formulate the Ginzburg criterion \cite{ginzburg1, amit2}, which usually tells us quantitatively when MFT is valid.  It is clear that the fluctuations become more and more pronounced as the temperature approaches the true critical point $T^{\ast}_c$ .  The Ginzburg criterion indicates in a semi-quantitative manner the temperature range where the distance from the SCM is important.

Therefore, the critical Ginzburg width \cite{ginzburg1, levanyuk} in the vicinity of the critical temperature is then given by
\begin{equation}
 \Delta t_G  = \Big|\frac{T^{\ast}_c - T_c}{T_c}\Big| = \frac{\Omega_{dc}}{a_0T_c}.
\end{equation}
In the last formula, ${\Omega_d}_c = \Omega_d(T^{\ast}_c)$ is the solution of the following self-consistent equation (obtained in the same spirit as Eq. (19) in \cite{roger}):
\begin{equation}
\Big(\frac{d}{2} - 2 \Big)\ln\Big[|\epsilon| + \frac{{\Omega_d}_c}{a_0T_c}\Big] = \ln\Big[ \frac{{a^2_0T^2_c{\Omega}_d}_c}{K_d(a_0T_c - {\Omega_d}_c)^2}\Big]
\end{equation}
Here, $T^{\ast}_c$ also determines the temperature under which the description of fluctuations goes beyond the independence of the MFA modes due to precise agreement. $\Delta t_G$ corresponds to the width of the critical region about the real transition temperature for which both the Landau and the GL theories are valid.  Further by taking into account the effective GLW Hamiltonian, this quantity also defines the domain of validity of the classical critical behavior \cite{goldenfeld}. The classical description fails for $|\Delta t| \equiv |T/T^{\ast}_c - 1| \ll |\Delta t_G|$. This finding is corroborated by the Ginzburg criterion and can be interpreted as \cite{amit, ginzburg1}. Thus, the Ginzburg criterion allows us to restore some credibility to the MFT in those cases. Eq. (25) shows that the anharmonic fluctuation $\Omega_d \rightarrow 0$ when $V \rightarrow \infty$ involving $a^{\ast}(T) \rightarrow a(T)$. In the thermodynamic limit, the renormalized quadratic coefficient $a^{\ast}(T)$ is the same as in the LT. It is well established that power-law predictions of the LT are correct in the thermodynamic limit, $T^{\ast}_c \rightarrow T_c$ when $V \rightarrow \infty$.

As it is well established (for $V \neq \infty)$, no Gaussian approximation even renormalized in some way can describe precisely critical phenomena near the transition point. Also, within this approach, the effects of spatial fluctuations are strongly dependent on lattice dimensionality and we can now appreciate how fluctuations and correlations modify the  macroscopic thermodynamic properties.

From the thermodynamic definition, the inverse renormalized susceptibility is given by the following analytic expression:
\begin{equation}
{\chi^{\ast}}^{-1} = a_0(T-T_c) + \Omega_d.
\end{equation}
Taking a look at the influence of spatial fluctuations on the stability of the renormalized mean-field analysis, the importance of fluctuations seems to be evident and the divergence of the susceptibility is related to the system size $L$ and dimensionality $d$.

The correlation length gives information about the distance for which the order parameter $\phi$ varies in the space. In order to calculate the correlation range, we first evaluate $\partial^{2} G^{\ast}(q)/\partial q^{2}$ where $G^{\ast}(q)$ is the renormalized $q$-mode autocorrelation function. Taking into account Eq. (20), the function $G^{\ast}(q)$  is obtained as
\begin{eqnarray}
G^{\ast}(\textbf{q}) = \langle{\vert\phi(\textbf{q})\vert}^2\rangle = {\left[2\beta \bigg(a(T) + \Omega_d + q^2\bigg)\right]}^{-1}.
\end{eqnarray}
Accordingly, as $T\longrightarrow T^{\ast}_c$ within the limit $q\longrightarrow 0$, it diverges according to $G^{\ast}(\textbf{q}, T = T^{\ast}_c) \sim q^{\eta - 2}$, with a small positive value $\eta$ and the critical exponent $\eta = 0$.  Therefore,  $\bar{G}^{-1}(q)$ appears as the reduced inverse q-mode autocorrelation function. According to Eq. (32), the renormalized correlation length is obtained as
\begin{eqnarray}
&&\xi^{\star}_d (T) = {\left[-\lim_{q \to 0}{G^{\star}}^{-1} (q){\left(\frac{\partial^2 G^{\star}(q)}{\partial q^2}\right)}\right]}^{1/2} \nonumber\\
&& = \left\{  \begin{array}{ll}\vspace{0,5cm}
\xi_0^+ T_c^{1/2}\Big[T-T^{\star}_{c} + \frac{1}{a_0}(\Omega_d - {\Omega_d}_c)\Big]^{-1/2} ,  & \textrm{for $T>T^{\star}_{c}$}\\
\frac{1}{2}\xi_0^- T_c^{1/2}\Big[T^{\star}_{c}-T + \frac{1}{a_0}({\Omega_d}_c - \Omega_d)\Big]^{-1/2},  & \textrm{for $T<T^{\star}_{c}$}
  \end{array}  \right.
\end{eqnarray}
which generally sets the characteristic length scale of fluctuations. The universal exponents and amplitude ratios are again recovered from this equation with regard to the critical temperature $T^{\ast}_{c}$.  Hence, the corresponding critical exponents are $\nu = \nu'$ = 1/2. The temperature dependence of the renormalized correlation length $\xi^{\ast}_d(T)$ based on the above formula depends on the dimensionality $d$ and the size $L$ of the physical space and behaves differently in
 function
of them.

For finite-size systems (in d = 2,3) it has been recognized \cite{fisher2, fisherbarber} that the system size L "scales" with the correlation length of the bulk system. Thus it is convenient to define a reduced length of the system:
\begin{eqnarray}
l^{\ast} = L/\xi^{\ast} = l_{0}T_c^{-1/2}\Big[T^{\ast}_{c}-T + \frac{1}{a_0}({\Omega_d}_c - \Omega_d)\Big]^{1/2},  & \textrm{for $T<T^{\star}_{c}$}.
\end{eqnarray}
The value of $l_{0} \sim \xi_0^{-1}$ depends only on the nature of the substance under study. In fact both $l^{\ast}$ and $l_{0}$ are dimensionless, and they are the ratio of the real thickness of the system to certain characteristic length. Indeed, if $l^{\ast} \gg 1$, no significant finite-size effects should be observed. On the other hand, for $l^{\ast} \leq 1$, the system size will cut-off long-distance correlations so that an appreciable finite-size rounding of critical-point singularities is to be expected. This result is not surprising considering the physical meaning of the correlation length which can be regarded as an indication of the influence range of the boundary condition. One should notice that what plays a role is not the real length $L$ of the system but the reduce one $l^{\ast}$ which depends also on the deviation from the bulk critical point. Details will be discussed later in the section reserved for the applications. This shows that if the reduced length of the system is too short, either due to the small size of the system or due to its closeness to the bulk critical point, the influence of the boundary will strongly dominate. As $l^{\ast}$ becomes large, the renormalized $H_{0}$ becomes independent of details of the system at the atomic level. This leads to an explanation of the universality of critical behavior for different kinds of systems at the atomic level. Liquid-gas transitions, magnetic transitions, alloys transitions, etc., all show the same critical exponents experimentally; theoretically this can be understood from the hypothesis that the  same "fixed point" interaction describes all these systems.

In order to illustrate the meaning of the correlation length, it is perhaps worth calculating the space-dependent correlation function $G^{\ast}_d(X)$ defined by the Fourier transform of the q-mode autocorrelation function as
\begin{equation}
G^{\ast}_d(X) = L^{-d/2}\int\exp(iqX)G^{\ast}(q)d^{d}q ,
\end{equation}
which can also be considered here in the sense of the entanglement in the model system.
The propagator $G^{\ast}_d(X)$ should always go to zero at large distances, so that a measure of order in the system is the long-length behaviour of the propagator.  If the propagator goes to zero then the system can only have short-range order. If the propagator goes to a non-zero constant at large distances, then we must have a non-zero order parameter, and the system has long-range order.

The integral (35) is well behaved at small $q$. At large $q$ without the exponential the integral would only be convergent for $d$ = 1. For $d$ = 2 there would be a log divergence, while for higher $d$ there would be an algebraic divergence. Clearly this divergence is cut off by the exponential factor. As the exponential goes to 1 when $X$ goes to zero, the correlation function should "diverge" at small distances. In fact the divergence will be cut off by the discretization length $\Lambda$ (discreteness of underline lattice). The choice of scale depends on the situation we want to study. First of all, in a numerical attempt to describe experimental data, it is clear that for most practical cases the cutoff $\Lambda$ must not be chosen too large, since this would require a knowledge of the lattice dispersion way beyond the parabolic approximation corresponding to the simple gradient correction in real space. However, for the study of phase transitions, this is not a severe restriction since the anomalies appearing in such transitions result from long range correlations, i.e. the behavior of the system at small $q$-vectors. For large distances $X > \xi$ , the exponential is oscillating much faster than any other variations, and the correlation function should fall rapidly to zero (in fact exponentially). This behaviour is demonstrated if we explicitly calculate the integral for dimensions one, two and three.

Using Eq. (35) and due to the fact that the space-dependent correlation function assumes various forms depending on the dimensionality of the physical space \cite{stanley}, $G^{\star}_d(X)$ satisfies
\begin{eqnarray}
G_d^{\ast}(X) =  \left\{  \begin{array}{ll}\vspace{0,5cm}
{\Big(\frac{x}{\xi^{\ast}}\Big)} \exp{\left[-{\Big(\frac{x}{\xi^{\ast}}\Big)}\right]},  & \textrm{for d = 1}\\\vspace{0,5cm}
 \frac{1}{2\pi}K_0\Big(\frac{\rho}{\xi^{\ast}}\Big),  & \textrm{for d = 2}\\
\Big(\frac{1}{4\pi}\frac{1}{r}\Big)\exp{\Big[-{\Big(\frac{r}{\xi^{\ast}}\Big)}\Big]} & \textrm{for d = 3},\\
  \end{array} \right.
\end{eqnarray}
where $\rho = (x^2 + y^2)^{1/2}$, and $r = (x^2 + y^2 + z^2)^{1/2}$.  $K_0(\textbf{r})$ is the modified Bessel function (that is logarithmic for small arguments and exponentially decaying for large arguments).

To keep the discussion self-contained, in the remainder of this subsection we review the predictions for the static critical exponents. First, we define the reduced temperature $\epsilon^{\ast} = (T - T^{\ast}_{c})/T_{c}$. The exponents $\alpha$, $\beta$, $\gamma$, $\eta$ and $\nu$ describe the  singular behavior of the  theory with strictly zero renormalized quadratic coefficient as $\epsilon^{\ast} \rightarrow 0$. For the specific heat, taking into account Eq. (23) one finds
\begin{equation}
C(T) \sim |\epsilon^{\ast}|^{-\alpha} + \textrm{less singular terms}
\end{equation}
$\eta$ and $\nu$ describe the behavior of the correlation length $\xi$, where
\begin{equation}
G_{\alpha\beta}(r) \equiv \langle\phi(r)_{\alpha}\phi(0)_{\beta}\rangle - \langle\phi_{\alpha}\rangle \langle\phi_{\beta}\rangle
\end{equation}
and the exponent $\eta$ is defined through the behavior of the Fourier transform of the correlation function:
\begin{equation}
G_{\alpha\beta}(q \rightarrow 0) \sim q^{-2+\eta}.
\end{equation}
The correlation length  exponent $\nu$ is defined by
\begin{equation}
\xi \sim |\epsilon^{\ast}|^{-\nu}
\end{equation}
The exponent $\beta$ will be defined later. The last  exponent, $\delta,$ is related to the behavior of the system in a small magnetic field $h$ which explicitly breaks the $\mathcal{O}(4)$ symmetry. The six critical exponents defined above are related by  four scaling relations \cite{ma}.
\begin{equation}
\alpha = 2 -d\nu, \phantom{..} \alpha + \beta(1+ \delta) = 2 = \alpha + 2\beta + \gamma, \phantom{..} \gamma = \beta(\delta -1).
\end{equation}

%
\section{Renormalized $\phi^{4}$ theory}
%
All perturbative approaches are based on the division of the free energy into a Gaussian term and higher order perturbative terms. In fact, the coupling constant with $\phi^{4}$ model of the perturbation is not necessary small, so that the convergence of the perturbation expansion cannot be ensured. Thus some more effective approaches to the calculation are needed. For systems with boundaries, one should consider the influences of the boundaries on the thermal properties near the bulk critical point. In additions, the spatial distributions of the order parameter should be taken into account for finite-size systems. As is well known for finite-size system, however, the spatial distribution cannot be considered as uniform any longer due to the influence of the boundary though the condition of minimum free energy would prefer a smooth distribution.
%
\subsection{Continuous phase transitions}
%
%
Different thermodynamic phases are characterized by certain macroscopic, usually extensive state variables called order parameters; examples are the magnetization in ferromagnetic systems, polarization in ferroelectrics, and the macroscopically occupied ground-state wave function for superfluids and superconductors. We shall henceforth set our order parameter to vanish in the high temperature disordered phase, and to assume a finite value in the low-temperature ordered phase.  Landau's basic construction of a general mean-field description for phase transitions relies on an expansion of the free energy (density) in terms of the order parameter, naturally constrained by the symmetries of the physical system under consideration \cite{tauber1}. For example, consider a scalar order parameter $\psi$ with discrete inversion or $Z_{2}$ symmetry that in the ordered phase may take either of two degenerate values $\psi_{\pm}= \pm|\psi_{0}|$ \cite{tauber2}. We shall see that the following generic expansion (with real and renormalized coefficients) indeed describes a continuous or second-order phase transition:
\begin{equation}
F^{\ast}_{L} = a^{\ast}(T)\psi^{2} + b^{\ast}(T)\psi^{4} + ... - h\psi,
\end{equation}
if the temperature-dependent parameter $a^{\ast}$ changes sign at $T^{\ast}_{c}$. For simplicity, and again in the spirit of a regular Taylor expansion, we let $a^{\ast}(T) = a_{0}(T - T^{\ast}_{c})$, where $T^{\ast}_{c}$ denotes the critical temperature from mean-field approach with renormalized $\phi^{4}$ model. The free energy is of almost the same as the Landau functional, except for the presence of the fluctuations-dependent in the quadratic and quartic coefficients, which contain the essential information about the microscopic nature of the system, its size and dimension. Our functional therefore describes a set of interacting, weakly-GLW-damped excitations.

 Stability requires that $b^{\ast} > 0$ (the size of the sample is such that the renormalized quartic term is a positive quantity); near the renormalized critical point we can simply evaluate $b^{\ast}$  at $T^{\ast}_c$. Details will be discussed shortly. Note that the external field h, thermodynamically conjugate to the order parameter, explicitly breaks the assumed $Z_{2}$ symmetry $\psi \rightarrow -\psi$. Minimizing the free energy with respect to $\psi$ then yields the thermodynamic ground state. Thus, from $\partial{F^{\ast}_{L}}/\partial\psi = 0$ we immediately infer the equation of state
\begin{equation}
h(T, \psi) = 2a^{\ast}\psi + 4b^{\ast}\psi^{3}
\end{equation}
and the minimization or stability condition reads $0  < \partial^{2}{F^{\ast}}_{L}/\partial\psi^{2}= 2a^{\ast} + 12b^{\ast}\psi^{2}$. At $T = T^{\ast}_{c}$, Eq. (43) reduces to the \emph{critical isotherm} $h(T^{\ast}_{c}, \psi) = 4b^{\ast}\psi^{3}$. For $a^{\ast}(T) > 0$, the \emph{spontaneous} order parameter at zero external field $h = 0$ vanishes; for $a^{\ast}(T) < 0$, one obtains $\psi_{\pm}= \pm|\psi_{0}|$, where
\begin{equation}
\psi_{0} = \sqrt{|a^{\ast}|/2b^{\ast}} = f(T)\cdot\phi_{0}
\end{equation}
 $\phi_{0} = \sqrt{|a|/2b}$ is the order parameter in LT. The behavior of the renormalized order parameter defines $\beta$:
\begin{equation}
\langle|\psi|\rangle \sim |\epsilon^{\ast}|^{\beta} \phantom{...} \textrm{for} \phantom{..}\epsilon^{\ast} < 0
\end{equation}
Taking into account Eqs. (12) and (13), the size dependence of $\psi$ enters in the amplitude prefactor $f(T)$ defined by
\begin{equation}
f(T) = \sqrt{\frac{|1 + \Omega_{d}/a(T)|}{1 + \Theta_{d}/b(T_c)}}.
\end{equation}
The SCM (taking into account both the dimension, the finite-size effects and the temperature dependence of $f(T)$) reveals a  competition between three scales of energies that are in competition: the thermal energy $a(T)\propto k_{B}T$ versus the energy resulting to anharmonic instabilities $\Omega_{d}$ on one hand, and  the anharmonic instabilities versus the harmonic instabilities on the other hand. These competitions determine both the existence of the transition, but also the adiabatic regime of fluctuations of the "pre-transition".

 As in many case, it is also possible to write the generalized GLW functional in a form which preserves certain transformations of $\psi(\textbf{r})$ like either of the three.
\begin{equation}
 \textrm{reversal}     \phantom{.......} \psi(\textbf{r}) \rightarrow -\psi(\textbf{r})
\end{equation}
\begin{equation}
 \textrm{change of phase} \phantom{.......} \psi(\textbf{r}) \rightarrow e^{i\theta}\psi(\textbf{r})
\end{equation}
\begin{equation}
\textrm{rotation}    \phantom{.......} \psi(\textbf{r}) \rightarrow U\psi(\textbf{r})
\end{equation}
(U is a rotation matrix). This holds for example for the Ising model, superfluid helium and the Heisenberg model respectively. The amount of $\psi$ is uniquely defined, but the sign, phase and direction of $\psi(\textbf{r})$, 
 are not defined. It depends on the history (preparation in an external field which removes the symmetry Eq. (48) or accidental fluctuations) of the system.

Note the emergence of characteristic power laws in the thermodynamic functions that describe the properties near the renormalized critical point located at $T = T^{\ast}_{c},$  $h = 0$.
Inserting Eq. (44) into the Landau free energy Eq. (42) one finds for $T < T^{\ast}_{c}$ and $h = 0$
\begin{equation}
F^{\ast}_{L}(\psi_{\pm}) = \frac{a^{\ast}}{2}\psi^{2}_{0} = -\frac{{a^{\ast}}^{2}}{4b^{\ast}}.
\end{equation}
Because of  our assumptions about $a^{\ast}$ and $b^{\ast}$,  the renormalized free energy  is proportional to  $(T^{\ast}_{c} - T)^2$.  This is characteristic of  all second-order transitions, and consequently for the specific heat
\begin{equation}
C^{\ast}_{h=0} = -L^{d}T\bigg(\frac{\partial^{2} F^{\ast}_{L}}{\partial T^{2}}\bigg)_{h=0},
\end{equation}
whereas by construction $F^{\ast}_{L}(0) =  0$ and $C^{\ast}_{h=0} = 0$ in the disordered phase. Thus, Landau's renormalized MFT also predicts a critical point discontinuity
\begin{equation}
\Delta{C^{\ast}}_{h=0} = L^{d}T^{\ast}_{c}\frac{a_0^2}{2b^{\ast}}\Bigg(1 + \frac{\partial\Omega_d}{\partial{T}}\Big|_{T = T^{\ast}_{c}} \Bigg)^2,
\end{equation}
for the specific heat.  Experimentally, one indeed observes singularities in thermodynamic observables and power laws at continuous phase transitions, but often with critical exponents that differ from the above mean-field predictions.  Indeed, the divergence of the order parameter susceptibility indicates violent fluctuations, inconsistent with any mean-field description that entirely neglects such fluctuations and correlations. Let us mention that the jump of the heat capacity was obtained because of the system volume was taken to infinity first, and after this the reduced temperature $\epsilon^{\ast}$ was set equal to zero.

Within the framework of this approach, taking into account the fact that $\frac{\partial\Omega_d}{\partial{T}}\Big|_{T = T^{\ast}_{c}} = 0$ and $b^{\ast}(T^{\ast}_c) = b(T_c)$, the anomalous part of the specific heat is really given by
\begin{equation}
\Delta{C^{\ast}}_{h=0} = L^{d}T_{c}\frac{a_0^2}{2b(T_c)}\Bigg(1 - \frac{\Omega_{dc}}{{a_0T_c}} \Bigg).
\end{equation}
Then, it can be concluded that (taking into account new fluctuating quantities) if the transition takes place at low temperatures, the anomalous part of the specific heat could be negligible or important compared to Debey's specific heat stemming from the acoustic phonon distribution \cite{roger}. Our approach suggests the first case where this anomalous part could be too small to be detected.  Eq. (53) is an indication of the decreased anomalous part of the specific heat as the anharmonic fluctuation of the system increases.

A close look at the heat capacity transition  in optimally doped YBCO samples  shows that  it  starts  several  degrees  above  $T_c$,  and presents a rather sharp peak, with an increasing slope  $(dC^{\ast}/dT)$ as  $T_c$ is  approached \cite{junod}. These are strong indications that  thermodynamical fluctuations are playing an important role in the transition.  We  can use Eq. (29) to estimate  the width of the critical region in a typical cuprate.

A different deviation from  mean field  behavior is also seen in the  transition of high $T_c$ cuprates, as shown  in \cite{junod}. In  addition to  some broadening observed above $T_c$ the shape of the main transition is modified. Instead of a jump, it looks more as a narrow  peak.  Qualitatively, it  reminds of the  specific heat peak seen at the transition of superfluid  Helium.

An even more radically different form of heat capacity transition is observed in $Bi_{2}Sr_{2}CaCu_{2}0_{8+ \delta}$. There is no more heat capacity jump at the transition, but rather a cusp.  The transition is better fitted  by a Bose-Einstein condensation than by a BCS one \cite{deutscher}. More details will be discussed shortly.

%
\section{Examples of applications}
%
%
In order to test the theory of critical phenomena it is important to have accurate experiments on well characterized systems, very close to the critical point. Such experiments exist in magnetism, ferroelectric thin films and superconductors.
%
\subsection{Magnetism}
%
%
\qquad Magnetism is caused at the atomic level by unpaired electron with magnetic moments, and in a ferromagnet, a pair of nearby electron with moments aligned has a lower energy than if the moments are antialigned \cite{wilson3}. The Curie point of a ferromagnet will be used as a specific example of a critical point. Below the Curie temperature $T_c$, the ideal ferromagnet exhibits spontaneous magnetization ($\phi \neq$ 0) in the absence of an external field; the direction of the magnetization depends on the history of the magnet. Above the Curie temperature, there is no spontaneous magnetization. This ferromagnetism is observed in certain metals like iron, nickel and cobalt. Just below the Curie temperature the mean field magnetic susceptibility is observed to behave as
\begin{equation}
\chi_m = \frac{C}{{\vert T-T_c \vert}^{\gamma}},  \hspace{0.5cm} \gamma = 1
\end{equation}
that is the Curie-Weiss law. $C$ is the Curie constant \cite{buschow, coey}. Experimentally, we observe that $\gamma$ is about 4/3. On the other hand, the reduced magnetization $\bar{\phi} \equiv \phi/(N\mu_B)$ (that is also the local spin density) is observed to behave as
\begin{eqnarray}
{\bar{\phi}}_{LT}  \propto  \left\{ \begin{array}{ll}
0  & \textrm{$T >T_{c}$} \\
(T_c - T)^{\beta} & \textrm{$T <T_{c}$}
\end{array}\right.
\end{eqnarray}
i.e. the exponent $\beta$ is 1/2, which disagrees with the evidence, experimental and theoretical, that $\beta$ is about 1/3.

As we showed above, the approach that we propose preserves the mean-field critical exponents; all the critical mean-field exponents are suitable. Here we do not renormalize the critical exponent from the renormalization-group viewpoint, but we take into account the explicit fact that the mean-field critical temperature $T_c$ is  in reality  a characteristic scale of temperature linked to the thermal fluctuations rather than to the transition temperature.

%
\subsubsection{Renormalized Weiss theory}
%
%

Ferromagnetism and the Curie temperature were explained by Weiss in terms of a huge internal "molecular field" proportional to the magnetization. The theory is applicable both to localized and delocalized electrons. No such magnetic field really exists, but it is a useful way of approximating the effect of the interatomic Coulomb interaction in quantum mechanics. When the distance between magnetic moments is small, the Pauli exclusion principle, which states that two identical fermions may not have the same quantum states, results in interaction between magnetic moments. Heisenberg introduced a model to describe this exchange interaction on microscopic scale. The Heisenberg exchange Hamiltonian may be written in the form
\begin{equation}
H_{\textrm{exch}} = -2\sum_{i < j}J_{ij}S_{i}\cdot S_{j}
\end{equation}
the summation extends over all magnetic moment pairs in the crystal lattice. For
positive values of the exchange constant $J_{ij}$ one finds parallel alignment else antiparallel. Ferromagnetism is observed for positive exchange interactions below a critical temperature.

The exchange interaction can be regarded as effective field acting on the moments. This field is produced by the surrounding magnetic moments and called here "renormalized molecular field". As the size of the surrounding moments is proportional to the magnetization, the renormalized molecular field $H^{\ast}_{m}$ is written as
\begin{equation}
 H^{\ast}_{m} = N^{\ast}_W M ,  \phantom{......} N^{\star}_W = N_W(1 - \frac{\Omega_{dc}}{a_{0}T_{c}})
\end{equation}
with $N^{\ast}_W$ the renormalized Weiss-field coefficient. It is necessary to note that $N_W$ was already introduced in the early $20^{th}$ century long before the development of quantum physics \cite{buschow, coey}. Within the framework of the Weiss theory, one postulates the existence of a mean field by a phenomenologic approach. The structure of parameter $N^{\ast}_W$, in such a theory, is not specified. We propose within the framework of this theory to make a microscopic justification of it;  we place ourselves within the framework of the "localized magnetism" where the elementary magnetic components are localized on each site of the crystal lattice. The anharmonic fluctuations tend to generate mechanisms which are opposed to the local order leading to a frustrated internal magnetism. The  Weiss coefficient  now takes into account both the dimension of the system, its size and critical fluctuation and  the total magnetic field experienced by a magnetic material is thus the sum of the externally applied field $H_{0}$ and the internal field,
\begin{equation}
H = H_{0} + H^{\ast}_{m},
\end{equation}
and Eq. (54) needs to be rewritten as
\begin{eqnarray}
\chi^{\star}_m = \frac{C}{{\vert T-T^{\ast}_c \vert}^{\gamma}}, & \textrm{where} \phantom{..} \gamma = 1 & \textrm{and} \phantom{..} T^{\star}_c = T_c - \frac{1}{a_0}{\Omega_{dc}}.
\end{eqnarray}
$T^{\ast}_c$ is the renormalized critical temperature and $T_c$ would be the critical temperature in absence of the correction term.

For most materials, the phase transition from the paramagnetic state to the ferromagnetic state is found to be of second order. This means that the temperature dependence of the first derivative of the free energy is continuous and the second derivative of the free energy is discontinuous. Within a  "renormalized molecular field" model we introduce an additional dimension $d$-dependent parameter
\begin{equation}
T^{\ast}_{c} = f(\nu_{d})\cdot T_{c}.
\end{equation}
The size dependence of $T^{\ast}_{c}$ enters in the amplitude prefactor $f(\nu_{d})$ defined by
\begin{equation}
f(\nu_{d}) = 1 - \nu_{d},  \phantom{..} \textrm{where} \hspace{0.5cm}  \nu_{d} = \frac{\Omega_{dc}}{a_0T_c}.
\end{equation}

%
\subsubsection{The localized magnetism}
%
%

 The (localized) ferromagnetism is characterized by the short-range exchange interactions \cite{kittel}. In the Ising-like model, it is shown for instance that $\xi_0 \sim d_0$ ($d_0$ is the lattice periodicity ), which is in the order of some few {\AA} \cite{ zinnjustin1}. Indeed, the critical width $\Delta t_{G} \sim 10^{-1}$ is appreciable and implies a degree of resolution experimentally accessible and lets us predict strong corrections to the MFT and SGA \cite{fisher2}. Indeed, the discrepancy between the mean-field results and experiment though the disagreement between the critical exponents for different dimensionalities suggests that the mean-field results are too universal and suggests some essential dependence of the critical point on $d$ \cite{goldenfeld}. To understand the mechanism by which $\xi_0$ sets the characteristic length scale of fluctuations we have to consider the scattering amplitude.
 %
\subsection{Ferroelectric thin films}
%
%
\qquad Finite-size effects in dipolar (magnetic or ferroelectric) ultrathin films is a topic of growing interest for both technological and fundamental reasons, (see, e.g., references [39--47]). One particularly important issue in these low-dimensional systems is the dependency of their Curie temperature (that is, the highest temperature at which a spontaneous polarization  exists) on the film's thickness. The technological relevance of knowing such dependency partly stems from the possibility of designing improved devices since many properties fundamentally depend on the Curie temperature. Finite-size scaling theory predicts that this critical  temperature shifts to lower temperatures than that of the bulk when one or more of the material's dimension(s) is reduced to an atomic size \cite{fisherbarber}. The  critical temperature $T_{c}(L)$ at which the finite system begins to have nonzero average order parameter is given by \cite{yang}
\begin{equation}
T_{c}(L) = T_{c} - \frac{\pi^{2}}{a_0L^{2}}.
\end{equation}
The existence of a minimum reduced length $L_{\textrm{min}}$ \cite{batra, tilley} is an indication of the first shift of critical point for finite-size system with respect to the bulk one. If the length $L$ of the system goes to infinity (thus the system becomes unlimited), $T_{c}(L)$ in the last equation returns to the bulk value $T_{c}$. But for small $L$ the first shift of critical temperature will be important.

Crucial characteristics  of  an  important  class  of  materials  are  currently unknown. In particular, one may wonder if relation among Eq. (62) better describes the transition temperatures of ferroelectric thin films. In case that Eq. (62) holds  in  ferroelectric  films  (for  any  thickness  or "only" above some critical thickness), one may also wonder if the mean-field critical temperature, $T_{c}$, is consistent with a given universality class since there has been an intense debate for many years on whether ferroelectrics belong to some universality classes or not. This paucity of knowledge may arise from the fact that many effects can affect the intrinsic $T_{c}(L)$-versus-$L$ curve in ferroelectric thin films. Examples of such effects are the increasing importance of depolarizing fields when decreasing the film's thickness \cite{junquera} or the increasing release of the strain arising from the substrate for thicker films.\cite{canedy}.

Taking into account Eq. (34) recognizing that the system size L "scales" with the correlation length of the bulk system specially in the ordered phase,  Eq. (62) is rewritten as
\begin{equation}
T^{\ast}_{c}(l_{0}) = T^{\ast}_{c}\Big[1 - \bigg(\frac{\pi}{l_{0}}\bigg)^{2}\Big].
\end{equation}
If the length $l_{0}$ of the system goes to infinity (thus the system becomes unlimited), $T^{\ast}_{c}(l_{0})$ in the last equation returns to the real bulk value $T^{\ast}_{c}$. But for small $l_{0}$ the second shift (due to finite-size effects) of critical temperature will be important. Specially, if $l_{0} \rightarrow \pi$, $T^{\ast}_{c}(l_{0}) \rightarrow 0$, so that the limited system will stay in the disordered phase until the temperature becomes absolute zero (It should be kept in mind that $T^{\ast}_{c} \neq 0$ for  2D- and 3D-system \cite{roger}. Fluctuations in two dimensions are not expected to destroy the ordering as in the case of an isotropic Heisenberg ferromagnet because of the anisotropy of the coupling.). Thus there exists a lower bound for the size of the system above which there may exist phase transition.

A deviation of the $T^{\ast}_{c}(L)$-versus-L was experimentally found in Refs \cite{ambrose, zhang} for magnetic films below a critical thickness (note that such deviation carries important information, such as, e.g., the possible predominant  role  of  fluctuations and surface on the transition temperature of ultrathin films).

The traditional wisdom for ferroelectric thin films has been that for film thicknesses
smaller than 100 $nm$ the depolarization field will destroy any switchable  polarization making  small  particles or thin films non-ferroelectric \cite{batra, tilley}. The general  prediction was that the transition temperature $T_{c}$ will decrease with decreasing size and ferroelectricity will vanish below a minimum critical thickness.

Recent  thin  film  experiments \cite{bune} have however shown that switchable ferroelectric films can be made down to 0.9 $nm$ for a crystalline Langmuir-Blodgett deposited  random copolymer of vinylidene fluoride with trifluoroethylene, (PVDF-TrF70:30) on graphite. The minimum thickness is just two mono-layers. The critical size for small spherical lead zirconate-titanate (PZT) particles was calculated to be 25 {\AA}\cite{yamumoto}. Similarly it has been reported that switchable ferroelectric films can be made of PZT down to 3 or 4 $nm$ \cite{tybell} in disagreement with earlier theories but in agreement with recent theoretical calculations. This shows that finite-size effects impose no practical limitation  on thin  film ferroelectric memory capacitors though for some designs tunneling currents may become too large. The lateral width of the memory element as well does not  represent a limitation at present as no change in  the coercive field has been observed when the lateral size of the PZT cell was decreased  from 1 $\mu$ to 0.1 $\mu$. The voltage necessary for polarization reversal  in  such  a  100 $nm\phantom{.}\times$  100 $nm\phantom{.}\times$ 100 $nm$ cell is in the technologically accessible range of 5 $V$ \cite{gonzalo}.  The corresponding hysteresis curve has been  measured via the atomic force microscope (AFM) in the piezoelectric mode \cite{alexe}.

Taking into account Eq. (16), Eq. (63) can be put in the following form
\begin{equation}
T^{\ast}_{c}(l_{0}) = T_{c}\Big(1 - \frac{\Omega_{dc}}{a_{0}T_{c}}\Big)\Big[1 - \bigg(\frac{\pi}{l_{0}}\bigg)^{2}\Big].
\end{equation}
With regard to this last equation, a competition is to be envisaged between various types of fluctuations (thermal fluctuations and fluctuations due to finite-size effects) and its consideration in the self-consistent theory can show that finite-size effects impose no practical limitation  on thin  film ferroelectric, in agreement with recent theoretical calculations and certain effects observed by now \cite{gonzalo}.

It should be mentioned that finite-size effects such as depressions of $T_{c}$, and reductions in  $P_{S}$ (the spontaneous alignment of dipoles in the ferroelectrics),  have indeed been observed in some nano-crystals as small as 250 {\AA} in diameter \cite{ishikawa}. It is however known that at such sizes this is not an electrostatic phenomenon but seems to be due to surface strains or to inhomogeneity effects \cite{tanaka}. One should also stress that the local polarization $P$ as a function of the depth $z$ may actually increase as the surface is approached as in the case of PZT \cite{almahmoud} leading to an increase in $T_{c}$,  as the thickness $d$ decreases. Alternatively it may decrease resulting in a depression of $T_{c}$, with decreasing $d$.  Both effects have been observed by now. This can be described by the fact that the extrapolation length can have either sign leading to an increase of a  decrease of $P$ at the surface \cite{gonzalo}.

 %
\subsection{Superconductors}
%
%
\qquad In renormalized GLW theory the macroscopic wave function of the superconducting state
\begin{equation}
\psi(r) = \psi_0(r) exp[-i\varphi(r)]
\end{equation}
serves as the order parameter with the amplitude squared $|\psi_0|^2 = n_s$ being the density of the superconducting particles. We use here the macroscopic wave function that is a characteristics of the superfluid state of helium and of superconductivity.

Using the variation method, the two GLW renormalized equations
\begin{equation}
\frac{1}{2m}(-i\hbar\nabla + 2e\textbf{\textit{A}})^2\psi + a^{\ast}\psi + b^{\ast}|\psi|^2\psi = 0.
\end{equation}
\begin{equation}
j_s = \frac{ie\hbar}{m}(\psi^*\nabla\psi - \psi\nabla\psi^*) - \frac{4e^2}{m}|\psi|^2\textbf{\textit{A}}.
\end{equation}
are found. In the first term of Eq. (66), magnetic-field effects are included by making the usual replacement $\nabla \rightarrow \nabla + 2e\textbf{A}$. For the charge of the Cooper pairs here we have written -2$e$, although in the original formulation of the theory the electronic charge -$e$ was used.

If we consider a specimen with dimensions much greater than the penetration depth, the magnetic field vanishes inside the sample, and $\psi = \textrm{const}$. Therefore, only the last two terms of Eq. (66) are relevant, leading to $|\psi|^2 = - a^{\ast}/2b^{\ast}$. Of course, this result is identical to Eq. (44). Inserting this expression in Eq. (67), we obtain for the current density
\begin{equation}
j_s = \frac{2e^2}{m}\frac{|a^{\ast}|}{b^{\ast}}\textbf{\textit{A}}.
\end{equation}
Obviously, this expression for the supercurrent is identical to the second London equation. However the statistically appearing (thermal and voluminal) fluctuations result in an additional current density (or additional electrical conductivity).

Taking into account Eq. (12), the last equation can roughly be written in the form of two contributions:
\begin{equation}
j_s = j_{GL} + j_{\textrm{fluct}}.
\end{equation}
The first one, $j_{GL}$, just reproduces the London expression without fluctuations. The second term, i.e., the fluctuation part of the supercurrent
\begin{equation}
 j_{\textrm{fluct}} \simeq \frac{2e^2}{mb^{\ast}}\Omega_d\textbf{\textit{A}}
\end{equation}
has a more sophisticated nature. In order to carry out the $j_{\textrm{fluct}}$ term the anharmonic contributions in the GLW functional, originating from the fourth order term, have to be taken into account. Details will be discussed shortly. Consequently, the GLW renormalized expressions for the penetration depth $\lambda^{\ast}$ and the coherence length $\xi_{\textrm{GLW}}$ are, respectively,  given by
\begin{equation}
\lambda^{\ast} = \sqrt{\frac{mb^{\ast}}{2\mu_0e^2|a^{\ast}|}}
\end{equation}
\begin{equation}
\xi^{\ast}_{\textrm{GLW}} = \frac{\hbar}{\sqrt{2m|a^{\ast}|}}
\end{equation}
Both characteristic length scales, $\lambda^{\ast}$ and $\xi^{\ast}_{\textrm{GLW}}$, have the same dependence on $a^{\ast}$. Since their ratio $\kappa_{\textrm{GLW}}  =  \lambda^{\ast}/\xi^{\ast}_{\textrm{GLW}}$ is a function of $b^{\ast}$  only, it is temperature, dimension and size dependent,   given by
\begin{equation}
\kappa^{\ast}_{\textrm{GLW}} = \sqrt{\frac{m^2b^{\ast}}{\mu_0\hbar^2e^2}} = \kappa\sqrt{1 + \frac{\Theta_d}{b(T_c)}}.
\end{equation}
Practically, the mean-field parameter $\kappa$ is the so-called GL parameter  that allows a distinction between type I and type II superconductors.
This could not be true in high-temperature superconductors if we must take into account both the fluctuations and finite-size effects. The penetration depth, the coherence length, and the critical fields are intimately connected.

As we saw, $T_c$ is a characteristic scale of temperature related to thermal fluctuations and finite-size effects rather than the transition temperature, that also means that thermal fluctuations of the equilibrium  state also exist above the real critical temperature $T^{\ast}_c$. In the normal conducting state the deviation  from  equilibrium  can  lead  to  the  transient appearance of the superconducting state within certain regions, i. e., to the formation of "puddles" of Cooper pairs. These deviations from equilibrium are not stable, and they will disappear more or less quickly. The statistical appearance of Cooper pairs will become more and more rare the higher the temperature, since with increasing temperature the normal conducting state becomes more and more stable compared to the  superconducting state. Therefore, with increasing temperature, larger and larger deviations  from equilibrium are needed to generate the superconducting state. If we note further that the puddles of Cooper pairs represent perfectly conducting regions, we understand immediately that already above $T^{\ast}_c$ due to the fluctuations in the normal conducting state the statistically appearing puddles of Cooper pairs result in an  additional electrical conductivity, which must strongly increase on approaching $T^{\ast}_c$.

This influence of the thermal fluctuations can be clearly detected for a number of
superconductors. In \cite{li}, for example, it shows the transition curve of a bismuth film near
the mean-field critical temperature $T_c$. One sees there clearly  that  the  full  normal  resistance  is reached only at temperatures  considerably  above $T_c$ and the  electrical conductance  is  plotted  instead  of  the  resistance, the  additional  conductance $\sigma^{\ast}$ of the Cooper pair puddles, statistically appearing and vanishing again, is particularly clearly visible.

The additional conductance due to the Cooper pairs (that is  well  confirmed  by  experiment and that is due to the fluctuations.) can be calculated from the existing theories  of superconductivity in combination with the theory of fluctuations \cite{pressler, buckel}.

%
\subsubsection{Paraconductivity for $T > T^{\ast}_c$}
%
The fluctuations of the order parameter for $T > T^{\ast}_c$ can also contribute to the electrical conductivity, and the pretransitional rise in the conductivity is referred to as paraconductivity. The Fourier transform of the nonlocal electrical conductivity can be calculated with the aid of the Kubo formula
\begin{equation}
\sigma^{\ast}(k) = \frac{1}{2k_BT}\int_{-\infty}^{+\infty}dt\langle \hat{j}_k(t)\hat{j}_k(0)\rangle
\end{equation}
where the bracket implies a combined quantum mechanical and statistical average. Using the plane wave decomposition of $\psi(r)$ given in Eq. (3) and the definition of the GLW current operator given in Eq. (67) without magnetic-field effets, we have
\begin{equation}
j(r) = \frac{e\hbar}{m}\sum_{q, q'}(q + q')\psi_q\psi^{\ast}_{q'}\exp[i(q - q')\cdot{r}]
\end{equation}
Fourier transforming Eq. (75) in $\exp(iq\cdot{r})$ yields
\begin{equation}
j(k) = \frac{e\hbar}{m}\sum_{q}(2q + k)\psi^{\ast}_ {q}\psi_{q + k}
\end{equation}
Substituting Eq. (76) into Eq. (74) and limiting ourselves to the $k = 0$, we obtain
\begin{equation}
\sigma^{\ast}(0) = \frac{4e^2\hbar^2}{k_BTm^2}\int_{-\infty}^{+\infty}dt\sum_{q, q'}qq'\langle\vert\psi_q(t)\vert^2\vert\psi_{q'}(0)\vert^2\rangle.
\end{equation}
We assume terms with $q \neq q'$ are statistically independent and average to zero; thus
\begin{equation}
\sigma^{\ast}_{ij}(0) = \frac{4e^2\hbar^2}{k_BTm^2}\int_{-\infty}^{+\infty}dt\sum_{q}q_iq_j\vert\langle\psi^{\star}_q(t)\psi_{q}(0)\rangle\vert^2
\end{equation}
if we assume an exponential decay, the correlation function entering Eq. (78) may be written
\begin{equation}
\langle\psi^{\star}_q(t)\psi_{q}(0)\rangle = \frac{2m}{\hbar^2}\frac{k_BT}{q^2 + {\xi^{\ast}}^{-2}_{\textrm{GLW}}}\exp(-t/\tau_q),
\end{equation}
which reduces to the Ornstein-Zernicke form Eq. (32) for $t = 0$ in the case of superconductors. The relaxation time, $\tau_q$ will be calculated from the Landau-Khalatnikov model. Note that while we had an equation of motion involving the phase of the order parameter, we did not introduce an equation of motion for the magnitude, $\vert\psi\vert$. The time scale for achieving phase variations is necessarily slow (hydrodynamic) whereas $\vert\psi\vert$ may change rapidly (by the conversion of superconductor into normal conductor and vice versa). An exception is near the second order phase transition where the free energy becomes "soft" with respect to a variation of the magnitude of the order parameter and the relaxational dynamics slows down. This intuitive idea was quantified by Landau and Khalatnikov (1954) in their discussion of relaxation phenomena in superfluid $^4He$ near the lambda point; they made the ansatz \cite{ketterson}
\begin{equation}
\frac{\partial{\psi_q}}{\partial{t}} = -\Upsilon\frac{\hbar^2}{2m}\big(q^2 + {\xi^{\ast}}^{-2}_{\textrm{GLW}}\big)\psi_q,
\end{equation}
where $\Upsilon$ is a rate constant ($\Upsilon > 0$). The solution of Eq. (80) is given by
\begin{equation}
\psi_q(t) = \psi_q(0)\exp(-t/\tau_q), \phantom{...} \textrm{where} \phantom{...} \tau_q = \frac{2m}{\Upsilon\hbar^2}\frac{1}{q^2 + {\xi^{\ast}}^{-2}_{\textrm{GLW}}}.
\end{equation}
 Thermodynamic systems have characteristic microscopic (rapid) relaxation times and the
relaxation of the vast majority of their internal degrees of freedom proceeds on these time scales. There are two important exceptions: (i) modes involving degrees of freedom for which  conservation laws exist; and (ii) additional modes involving a broken symmetry of the system. Additional equations of motion exist when the system
spontaneously breaks some symmetry. Taking into account Eq. (81)  and inserting Eq. (79) into the k = 0 limit of Eq. (78), carrying out the integration over time, and writing $q^2_i = \frac{1}{d}q^2$, we obtain
\begin{equation}
\sigma^{\ast}(0) = \frac{16}{d}k_BT\frac{e^2m}{\Upsilon\hbar^4(2\pi)^d}\int\phantom{.}d^dq\frac{q^{2}}{(q^2 + {\xi^{\ast}}^{-2}_{\textrm{GLW}})^3}.
 \end{equation}
The integration over negative time is accomplished by replacing $\exp(-t/\tau_q)$ with $\exp(-\vert{t}\vert/\tau_q)$.

Inserting the expression for $\Upsilon$ obtained within the framework of time-dependent GLW renormalized theory in the vicinity of the renormalized critical temperature \cite{pressler, ketterson}
\begin{equation}
a_0\Upsilon = \frac{8k_B}{\hbar\pi},
\end{equation}

and taking into account the dependence of the correlation length $\xi^{\ast}_{\textrm{GLW}}$ on the temperature using Eqs. (12), (25) and (72), the singular component expression of  the  additional conductance is obtained as
\begin{equation}
\sigma^{\ast}_d(T) \simeq \aleph_d\frac{e^2\xi^{2-d}_{GLW}(0)}{\hbar}\Big(1 - \frac{\Omega_{dc}}{a_0T_c} \Big)\Big(\epsilon + \frac{\Omega_d(T)}{a_0T_c} \Big)^{\frac{d}{2}-2}
\end{equation}
where $\aleph_d = \frac{2^{1-d}\pi^{1-d/2}}{d\Gamma(d/2)}\int_{0}^{+\infty}\frac{x^{d+1}dx}{(1 + x^2)^3}$ is a constant integral which is dimension dependent and $\xi_{GLW}(0) = \frac{\hbar}{\sqrt{2ma_0T_c}}$. In low dimensions $d < 4$, we set $x = q\xi$  to render the fluctuation integral, which is UV-finite, dimensionless.

In Eq. (84) we have indicated the additional conductance for d-dimensional samples.

$\bullet$ In the case of three-dimensional superconductors, the thickness, the width, and the length of the sample are large compared to $\xi_{GLW}(0)$.
\begin{equation}
\sigma_{3D}^{\ast}(T) \simeq \frac{e^2}{32\hbar\xi_{GLW}(0)}\Big(1 - \frac{\Omega_{3c}}{a_0T_c} \Big)\Big(\epsilon + \frac{\Omega_{3D}(T)}{a_0T_c} \Big)^{-\frac{1}{2}}.
\end{equation}
$\bullet$ For a thin film with length $L$ small compared to $\xi_{GLW}(0)$, we may neglect the fluctuations along the film normal and the system becomes effectively 2D.
\begin{equation}
\sigma_{2D}^{\ast}(T) \simeq \frac{e^2}{16L\hbar}\Big(1 - \frac{\Omega_{2c}}{a_0T_c} \Big)\Big(\epsilon + \frac{\Omega_{2D}(T)}{a_0T_c} \Big)^{-1}.
\end{equation}
$\bullet$ In the case of one-dimensional superconductors,  the length of the sample is large compared to $\xi_{GLW}(0)$, the thickness $\ell$ and the width $\varrho$ are small compared to $\xi_{GLW}(0)$
\begin{equation}
\sigma_{1D}^{\ast}(T) \simeq
\frac{\pi{e^2}\xi_{GLW}(0)}{16\ell{\varrho}\hbar}\Big(1 - \frac{\Omega_{1c}}{a_0T_c} \Big)\Big(\epsilon + \frac{\Omega_{1D}(T)}{a_0T_c} \Big)^{-\frac{3}{2}}.
\end{equation}
$\Omega_{dc}$ is the the solution of the self-consistent Eq. (30) for one-, two- and three-dimensional sample. Qualitatively, we can easily understand that the sample dimensions must influence the magnitude of the fluctuations, since the Cooper pair density can vary only on a length scale of about $\xi^{\ast}_{GLW}$. More rapid spatial variations require relatively high energies and, hence, practically do not appear. Within a sample that is large in all three spatial directions, the Cooper pair density can vary spatially in all directions. All these possible configurations must be taken into account in the calculation of the  additional  conductance.  For  a  two-dimensional  sample,  along  the  shortest extension the Cooper pair density is always constant spatially. Hence, averaging over all possible spatial configurations of the Cooper pair density along this direction is not necessary. For a one-dimensional sample, averaging is unnecessary along both directions in which the sample is small compared to $\xi^{\ast}_{GLW}$. We see that the statistics is restricted because of the sample geometry. This results in various expressions for the additional conductance.

Experience shows that the transition curves of three-dimensional samples, say, of wires with a diameter large compared to $\xi^{\ast}_{GLW}$, are very sharp, i. e., the effects we have just discussed cannot be observed. The reason is not the absence of fluctuations, but rather the comparatively high residual conductance of the three-dimensional sample. The quantity $f(\Omega_{dc}) = \Big(1 - \frac{\Omega_{dc}}{a_0T_c} \Big)$ of Eqs. (85-87) is an indication of the decreased additional  conductance as the fluctuation effects of the system increase.

So far we have only discussed how the fluctuations affect the electrical conductance. However, if puddles of Cooper pairs appear statistically above $T^{\ast}_c$, this must be noticed also in other properties. We know that below $T^{\ast}_c$ a superconductor expels small magnetic fields out of its interior, i. e., it turns into an ideal diamagnet \cite{buckel}. We expect that, similar to the effect of the fluctuations on the conductance, some part of this diamagnetic property also appears above $T^{\ast}_c$.
The puddles of Cooper pairs should result in a characteristic temperature dependence of the diamagnetic behavior of the superconductor above $T^{\ast}_c$. Only a few hundredths of a degree away from $T^{\ast}_c$ the additional diamagnetism is already very small and corresponds to the expulsion of just a few flux quanta. However, it has been  possible  to  detect  this  effect  clearly \cite{gershenzon} by utilizing a superconducting quantum interferometer. The fluctuations should also lead to an increase of the specific heat $C$ already above $T^{\ast}_c$. This effect could also be experimentally demonstrated \cite{buckel, kreislerAJ}.

 %
\subsubsection{Conventional and unconventional superconductors}
%

The considerable success of the mean-field theory in conventional superconductors originates from the low value ($\Delta t_G = \Omega_{dc}/a_0T_c \sim 10^{-14}$) of the critical Ginzburg width [1, 18, 45]. Accordingly, Eq. (16) obviously leads  to  $T^{\star}_c \simeq T_c,$   where $T_c$ is the mean field (BCS) value of the transition temperature. This implies that it is impossible to detect with actual experimental precision the deviations from the mean-field theory. Indeed, one observes that the specific heat does not present any divergence, but  rather a compatible jump with the predictions of Eq. (3) in \cite{roger}. Practically, critical phenomena should not be observable in usual conventional superconductors, which could not be true in high-temperature superconductors and particularly high-$T_c$ superconductors \cite{bennemann, schrieffer}.

In compounds of the BiSrCaCuO, LaSrCaCuO and TlSrCaCuO type, estimations around $\xi_0 \sim $ 1 nm were reported, the paired electrons of the Cooper pair (the underlying unit for superconductors) are forced apart by their Coulomb repulsion, resulting in a relatively large separation of $\xi_0 \approx 10^{3}$ {\AA} \cite{bennemann, roger}. These values have to approximately equal the size of the units that undergo ordering at the phase transition. This can question the validity of the mean-field theory for superconductors. But, we think that such a distance is related to the existence of large fluctuations and finite-size effects present in those compounds. The increase of the critical temperature in non-conventional superconductors also means that the bond energy of the Cooper pair increases at the same rate. When the energy bond is higher, the bond is mostly confined in space and $\xi_0$ then reduces at the same level. It is for instance the case in $La_{2 - x}Sr_{x}CuO_4$ where $\Delta t_G$ can reach the value $10^{-1}$, and in $YBa_2Cu_3O_{7-y}$ compounds where $\Delta t_G$ can reach the values typically in the range $10^{-2}$ to $10^{-1}$ \cite{roger}. These values seem to be accessible in experiment and suggest principal deviations compared to the mean-field theory observable. Also, the newer ceramic high-temperature superconductors have a much smaller coherence length of $\xi_0$, which can indeed help to show some effects of fluctuations \cite{poole, owens, caruta}.

 %
\section{Conclusion}
%
%
We have described the fundamentals of renormalized $\phi^4$ theory on the basis of a GLW calculations,  combined with an efficient no perturbative technique, that takes into account both dimension, size and microscopic details of the system, and which leads to critical behavior, strongly deviating from the classical MFA far from the thermodynamic limit.  Within this more rigorous approach the effects of fluctuations are examined in more rigorous detail, and we are able to establish the insufficiencies of the MFT, and also estimate the width of the critical region where corrections to MFT are important.  We find a consistent interpretation  relating the dimension, the size of the system and the spatial fluctuations which can give an interpretation  to  the  degeneracy of energy levels. We have also calculated asymptotic expressions of thermodynamic observables as a function of temperature.

This theory does not present any renormalization based method in the sense of a renormalization group approach but rather a SCM improving on LT. The SCM is a strategy for dealing with problems involving many length scales. The  strategy is to tackle the problem in steps, one step for each length scale. In the case of critical phenomena, the problem, technically, is to carry out statistical averages over thermal fluctuations on all size scales. The SCM is to integrate out the fluctuations in sequence, starting with fluctuations on an atomic scale and then moving to successively larger scales until fluctuations on all scales have been averaged out. The integration of this deviation in the self-consistency on the entire spectrum of the lattice vibration, conducted to a very improved self consistent energy. This improvement of the self consistent problem by the processed harmonic and anharmonic fluctuations goes for example to reflect itself on the thermodynamic and electronic parameters, resulting notably in a substantial reduction of the critical temperature.

The resulting self consistent problem of this approach has been solved analytically, what permitted us to extract an effective theory, with notably a mean field critical temperature renormalized by fluctuations. An interesting point with this renormalized critical temperature is its dependence on the quantity $\frac{\Omega_{dc}}{a_0}$ which according to the text, characterizes the importance of the fluctuations and is a function of both dimension and size of the system.

We have revealed the thickness dependence of the Curie temperature for various types of thin and ultrathin films.  In agreement with recent theoretical calculations it is found that the relation of Eq. (64) reproduces rather well such dependence for thickness down to $l_{0}$, in contrast with the usual finite-size scaling law  that breaks down below $L$ and that does not take into account both the fluctuations and the microscopic details of the system. We are confident  that the present work provides a deeper knowledge of nanoscience, phase transitions, and critical behaviors in dipolar systems. Furthermore, the parameters characterizing this approach depend strongly on the dimension and the size of the sample, but in different ways than the ones obtained in the MFA. Thus, this SCM seems to be appropriate for model calculations in a number of critical systems. For
this reason, analysis has also been performed on other systems such as non-conventional superconductors, localized magnetism and can be extended to all system, whose transitions indeed apparently belong to the same universality class, in order to lend a significant level of support to this theoretical approach.

By comparing the thermal energy and the energy resulting to structural instabilities, we introduced the report $\nu_{d} $, which could allow to estimate the quantum fluctuation importance. For large values of $\nu_d$, only a quantum description can give account of the dynamics of the system and we must use another approach that is the second step in this theory and that is to write down a general field theory for the order parameter, consistent with all symmetries of the underlying model. As we are dealing with a quantum transition, the field theory has to extend over space time, with the temporal fluctuations representing the sum over histories in the Feynman path integral approach.

Moreover, the SCM appears very suitable for providing a quantitative description in the crossover region as its application makes it possible to address the importance of fluctuations and thence to elucidate the basic physics of phase transitions and critical phenomana. Finally, we also expect that our method will reveal universal features of critical systems with finite size, which needs numerical studies and pronounced experimental confirmation.

\newpage
%

%
\end{document}